# Increasing brightness in multiphoton microscopy with low-repetition-rate, wavelength-tunable femtosecond fiber laser

Jakub Bogusławski,* Alicja Kwaśny, Dorota Stachowiak,
and Grzegorz Soboń

*Laser & Fiber Electronics Group, Faculty of Electronics, Photonics and Microsystems,
Wrocław University of Science and Technology, Wybrzeże Wyspiańskiego 27, 50-370 Wrocław, Poland*
*jakub.boguslawski@pwr.edu.pl

**Abstract:** Many experiments in biological and medical sciences currently use multiphoton microscopy as a core imaging technique. To date, solid-state lasers are most commonly used as excitation beam sources. However, the most demanding applications require precisely adjusted excitation laser parameters to enhance image quality. Still, the lag in developing easy-to-use laser sources with tunable output parameters makes it challenging. Here, we show that manipulating the temporal and spectral properties of the excitation beam can significantly improve the quality of images. We have developed a wavelength-tunable femtosecond fiber laser that operates within the 760 – 800 nm spectral range and produces ultrashort pulses (<70 fs) with a clean temporal profile and high pulse energy (~1 nJ). The repetition rate could be easily adjusted using an integrated pulse picker unit within the 1 – 25 MHz range and without strongly influencing other parameters of the generated pulses. We integrated the laser with a two-photon excited fluorescence (TPEF) scanning laser microscope and investigated the effect of tunable wavelength and reducing the pulse repetition rate on the quality of obtained images. Using our laser, we substantially improved the images' brightness and penetration depth of native fluorescence and stained samples compared with a standard fiber laser. Our results will contribute to developing imaging techniques using lower average laser power and broader use of tailored fiber-based sources.

## 1. Introduction

Multiphoton microscopy uses nonlinear interaction with the sample as a contrast mechanism and is becoming widespread in biomedical sciences. Since the first demonstration of TPEF microscopy [1], tremendous progress has been made. This includes the development of instrumentation [2,3], electronics [4], detectors [5], various applications [6], and powerful data processing and visualization techniques, such as fluorescence lifetime phasors [7,8]. The advancement of TPEF microscopy permitted cellular imaging several hundred micrometers deep in various tissues and became the method of choice for imaging thick samples and living organisms [9]. While TPEF is the primary signal source in multiphoton microscopy, three-photon excited fluorescence, second- and third-harmonic generation (SHG, THG) can also be used for imaging [10,11].

Many fluorescent dyes and proteins are available with identified two-photon absorption properties [12]. In some cases, this allows one to choose the one with the proper absorption spectrum to match the excitation wavelength of the laser. Such comfort does not exist in autofluorescence imaging; additionally, action cross-sections of native fluorophores are much lower than most common dyes [10]. Furthermore, during *in vivo* imaging of native fluorophores, one must adhere to safety standards, limiting permissible excitation power [13,14]. Nevertheless, endogenous fluorophores carry vital information about the structure of the tissue, its composition, and its metabolic state. Consequently, the most demanding applications, such as neurology or ophthalmology, require precisely adjusted excitation laser parameters to enhance excitation efficiency.

In two-photon excitation, the average number of photons that the fluorescing medium emits per second is [15]:

$$\langle n_f(t) \rangle \propto \delta_2(\lambda_{TPE}) \cdot \eta \cdot g^{(2)} \cdot \frac{P_{avr}^2}{\tau_p \cdot f_{rep}} \left( \frac{NA^2}{\pi \cdot h \cdot c \cdot \lambda_{TPE}} \right)^2,$$

where $\delta_2(\lambda_{TPE})$ is a two-photon absorption cross-section, and $\eta$ is quantum yield. The first parameter describes the ability of a molecule to absorb photons at a specific wavelength, and the second specifies the efficiency of converting the absorbed photons to emitted ones. Factor $g^{(2)}$ is related to the energy within a pulse (0.66 and 0.59 for Gaussian and $sech^2$ pulses, respectively); $P_{avr}$ is the average laser power, $\tau_p$ is pulse duration, $f_{rep}$ is pulse repetition rate, $NA$ is the numerical aperture of the objective, $h$ is Planck constant, $c$ is the speed of light, and $\lambda_{TPE}$ is excitation wavelength. As the relationship depends on the square of the average power, increasing the excitation power is the most straightforward way of increasing the number of emitted fluorescence photons; however, it is not always possible due to safety standards and the risk of sample damage. The equation shows that the number of fluorescence photons can be increased by manipulating the excitation pulse train's temporal and spectral properties without increasing the deposited energy per time (e.g., per pixel). Specifically, one should use as short pulses as possible and reduce the pulse repetition rate. Such a redistribution of the incident photons in time enhances nonlinear interaction without changing the linear effects (such as sample heating). Additionally, the excitation wavelength should match the absorption cross-section of the fluorophore of interest. A laser source with a very short pulse, adjustable pulse repetition rate, and tunable wavelength would be highly desirable for efficient two-photon excitation.

Femtosecond Ti:sapphire lasers are the most commonly used light sources in multiphoton microscopy. On the one hand, they offer high average power, very short pulse duration, and broad wavelength tunability within the 680 to 1080 nm range, covering an optimum range for autofluorescence imaging (700-800 nm [16]). On the other hand, Ti:sapphire lasers are large, expensive, and require installation on the optical table and active water cooling, creating a barrier for the broader use of multiphoton microscopy. Additionally, they operate at a relatively high pulse repetition rate (ca. 80 MHz), and adding an external pulse picker unit makes the system more complex, costly, and introduces losses [6]. Femtosecond fiber lasers are an exciting alternative due to their smaller footprint, efficient air-cooling, ease of use, and lower cost. They do not require specialized maintenance and offer ultrashort pulse duration and sufficient power for most biomedical applications. The most common central wavelengths are 1030 nm (Yb-doped), 920 nm (Nd-doped), and 780 nm (Er-doped with frequency doubling); the latter is especially attractive for exciting native fluorophores. However, the lack of wavelength and other parameters tunability hinders the most demanding applications, especially in autofluorescence imaging.

Here, we report a compact fiber-baser femtosecond laser source with an adjustable pulse repetition rate and wavelength-tunable within 760-800 nm, making it suitable for autofluorescence imaging. Within this range, the laser produced pulses shorter than 70 fs with energy >1 nJ. The repetition rate could be adjusted between 1 and 25 MHz. We test the laser in the target application of multiphoton microscopy and compare its performance with a standard, commercially available fiber laser. Adjustable repetition rate and tunable wavelength were key features that allowed us to increase image brightness significantly, or alternatively, to reduce necessary excitation power. Additionally, substantially reducing the pulse repetition rate improved the penetration depth in volumetric imaging of scattering samples. Our results show that the developed flexible laser source enables unprecedented imaging quality at low excitation power.

## 2. Experimental setup and methods

Figure 1 presents the experimental setup containing four major units: femtosecond fiber laser, SHG module, multiphoton microscope system, and data acquisition and processing. We compared the performance of our laser with a standard femtosecond fiber laser.

Femtosecond fiber laser was developed in all-polarization maintaining (all-PM) and all-fiber configuration. The system was seeded by a low-noise Er:fiber oscillator, described in detail in [17]. In short, the oscillator produced 400 fs pulses centered at 1561 nm with a 48.8 MHz repetition rate and 4 mW of average power. Using intracavity spectral filtering, we obtained reduced intensity (integrated relative intensity noise of 0.0142%, 10 Hz – 500 kHz) and phase noise of the oscillator (timing jitter of 1.71 ps, 3 Hz – 1 MHz frequency range). The oscillator's output was followed by a 5% tap isolator feeding a reference frequency signal through the photodiode to the pulse picker driver (AA Optoelectronic PPKAc150-B-35). In the next step, the signal was preamplified in an Er-doped fiber amplifier (nLight Liekki Er80-8/125-PM) counter-pumped by a 976 nm laser diode. We used an acousto-optic modulator (AA Optoelectronic MT150-IIR10) with the computer-controlled diver in synchronization with the reference pulse train to reduce the pulse repetition rate. Before the final amplification, the pulse was filtered using a 10 nm bandpass filter. The power amplifier section consisted of 1.5 m of Er-doped fiber (nLight Liekki Er80-4/125-PM) co-pumped by a 976 nm laser diode. In this section, spectral broadening occurs due to self-phase modulation. The pulse was compressed in a short piece of anomalous dispersion passive fiber (Nufern PM1550-XP). At this point, 35 fs pulses with a central wavelength of 1560 nm, broad optical spectrum, and 6.1 nJ energy were generated (Fig. S1). The fiber-optic part was placed in three 3D printed enclosures (dim. of 15×11×2.5 cm each) to ensure constant temperature and operation conditions. The oscillator was placed on a metal printed circuit board heating plate with uniformly distributed heating traces. The temperature was kept at 28°C. Two amplifiers were placed on a metal board with thermoelectric coolers stabilizing the temperature at 25°C.

Next, the fiber-optics part was connected through the FC/APC connector with the SHG module used for frequency doubling and wavelength tuning. Two aspheric lenses (Thorlabs C220-TME-C, L1, L2) were used to collimate and focus the beam on a nonlinear crystal. We used a 0.5 mm periodically poled lithium niobate crystal (MgO:PPLN, Covesion MSHG1550-0.5-0.5) with nine quasi-phased matching (QPM) periods (18.5 – 20.9 μm). Having a broad fundamental spectrum and relatively thick crystal allows us to convert different parts of the spectrum by translating the crystal in front of the focused laser beam (and choosing one of the QPM periods). The crystal temperature was kept at 30°C. The frequency-doubled beam was collimated using an aspheric lens (Thorlabs A260TM-B, L3), and the beam size was 1.7 mm ($1/e^2$ criterion, Fig. S2). Two filters were used to remove the unconverted fundamental beam and parasitic wavelengths (Thorlabs FELH0700, F1, and Thorlabs FESH0800, F2, respectively).

We used a self-built two-photon excited fluorescence microscope setup to test the laser in its target application. The gradient index filter (Thorlabs NDL-25C-2) was placed at the entrance to the microscope to adjust the average excitation power. Next, the collimated laser beam was directed to a pair of close-coupled galvanometric scanners (ScannerMAX Saturn-5). Then, the light passed through a scan lens (Thorlabs SL50-2P2) and tube lens (Thorlabs TTL200MP) to the microscope objective (Nikon N20X-PF). Fluorescence light was collected using the same objective in the epi configuration. A dichroic mirror (Semrock HC 705 LP) and bandpass filter (Thorlabs FESH0700, passband 400-700 nm) separated fluorescence from the residual excitation light. An additional lens (Thorlabs AC254-050-A-ML) focused light on the photomultiplier tube (Thorlabs PMT1001/M) connected to the data acquisition card (PCIe-6363). We developed a custom control system using NI LabVIEW, connected to galvanometric scanners, photomultiplier, and microscope sample stage, providing simple integration with the computer [18]. Fluorescence images were further processed using MATLAB, and operations included frame averaging, thresholding to remove background noise, and image plotting using a perceptually uniform colormap [19]. For volumetric imaging, the collected data were processed using Python with numpy and vedo packages. The .txt files with recorded layers were read, and the value of each pixel was averaged. Then, we construct voxels from pixels and construct a 3D array, with which we initialize a vedo Volume object. We apply a threshold to our Volume object and create an isosurface, later saved as an .obj file.

We compared the performance of the developed laser in multiphoton imaging with a standard, commercially available fiber laser. As a reference, we used a Menlo T-Light femtosecond fiber laser. The laser produced 90 fs pulses with a central wavelength of 1560 nm and a 100 MHz pulse repetition rate. The laser was frequency doubled using the same SHG module and 19.4 µm QPM period. As a result, the laser produced 80 fs pulses with a central wavelength of 780 nm and 0.4 nJ energy (see Fig. S3 for more detailed information).

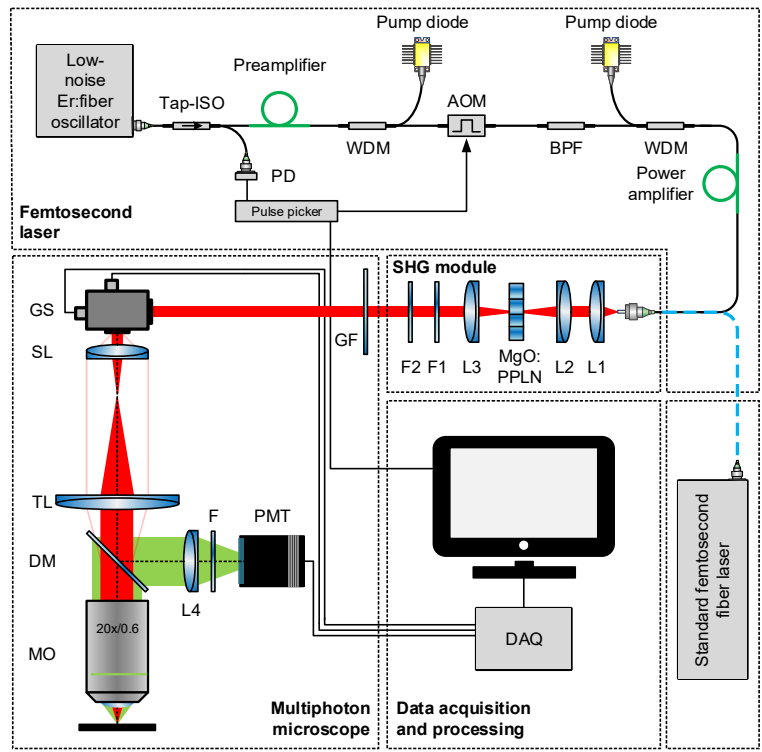

Fig. 1. Experimental setup of femtosecond fiber laser and multiphoton microscope. ISO – isolator, PD – photodiode, WDM – wavelength-division multiplexer, AOM – acousto-optic modulator, BPF – bandpass filter, L – lenses, F – filters, GF – gradient index filter, GS – galvanometric scanners, SL – scan lens, TL – tube lens, DM – dichroic mirror, MO – microscope objective, PMT – photomultiplier tube, DAQ – data acquisition card.

The lasers were characterized using an optical autocorrelator (APE PulseCheck), spectrometer (Broadcom AFBR-S20M2VN), digital oscilloscope (Siglent SDS2354X Plus), photodiode (Thorlabs DET10A/M), and power meter (Thorlabs PD100) with a semiconductor sensor (Thorlabs S120C) or thermal sensor (Thorlabs S401C).

## 3. Results

### 3.1 Laser characterization

Figure 2 presents the results of the characterization of SHG output. Seven (out of nine) quasi-phase matching periods were used, allowing for wavelength tuning. For each crystal setting, the pump powers of both amplifiers were adjusted to provide the shortest pulse; the pump power of the oscillator was kept constant. The laser produces very short femtosecond pulses for each crystal setting [53-70 fs after deconvolution assuming sech$^2$ pulse profile, as shown in Fig. 2(a)]. Corresponding optical spectra are shown in Fig. 2(b), demonstrating the tunability with a 761.5 – 798.5 nm range (center of the mass). For each central wavelength, the pulse energy is approx. 1 nJ or higher. The lowest energy of 0.97 nJ was recorded at the shortest wavelength of 761.5 nm, while the highest energy of 2.03 nJ was noted at the longest wavelength of 798.5 nm.

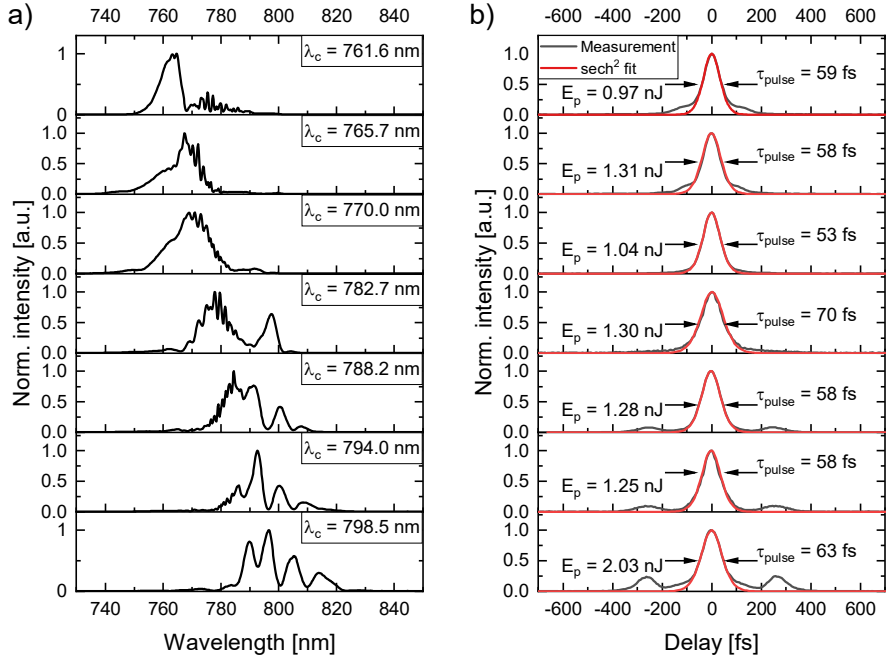

Fig. 2. Output characteristics of the laser after the SHG module recorded at 1 MHz pulse repetition rate. (a) Optical spectra were measured at seven QPM periods (18.5 – 20.3 μm, 0.3 μm difference between the periods) and (b) corresponding pulse autocorrelations; $\lambda_c$ indicates the center of the mass of the spectrum.

The integrated pulse picker unit provides the ability to adjust the pulse repetition rate without strongly influencing other parameters, such as pulse duration, optical spectra, and pulse energy. Figure 3(a) demonstrates the effect of pulse picker operation; the repetition rate could be adjusted within the 1 – 24.4 MHz range. For a given repetition rate, it was necessary to adjust the pump power of the power amplifier. Figures 3(b) and (c) show the autocorrelation function and corresponding optical spectra at four selected repetition rates with the example of the 19.1 μm QPM period. The pulse duration was maintained within the 50-58 fs range at this crystal setting, and similar behavior was observed for other periods. The output spectrum did not change substantially as well. The center of the mass was maintained within the 770.0 – 773.3 nm range. The pulse energy was 1.04 nJ, 0.97 nJ, 1.01 nJ, and 1.04 nJ for 24.4 MHz, 12.2 MHz, 5.4 MHz, and 1.0 MHz rates, respectively. This corresponded to 25.3 mW, 11.8 mW, 5.5 mW, and 1.04 mW output power for each setting.

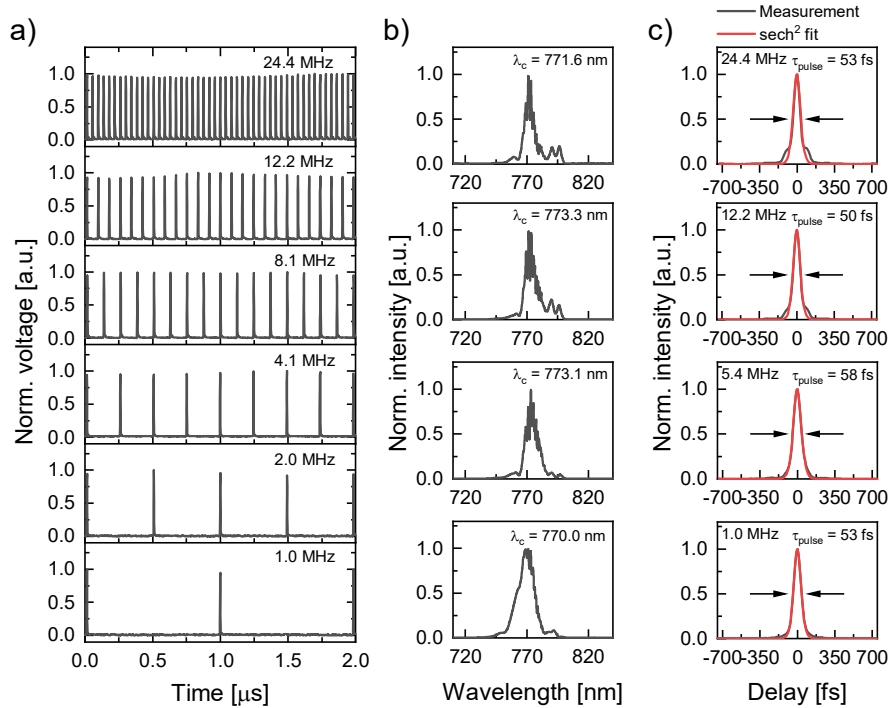

Fig. 3. Pulse repetition rate could be adjusted while marinating other parameters, like pulse duration, optical spectrum, and pulse energy. (a) Oscilloscope traces of output pulse after SHG module for selected pulse repetition rates with 1 – 24.4 MHz range. (b) Optical spectra of the output pulse at four selected repetition rates obtained at 19.1 μm QPM period and (c) corresponding pulse autocorrelations.

### 3.2   Effect of tunable wavelength

We first consider the effect of adjusting the central wavelength of the excitation beam in two scenarios: imaging stained sample and autofluorescence imaging. Figure 4(a) and (b) compare images of a convallaria majalis root transverse section at two excitation wavelengths, 761.6 nm and 788.2 nm. The sample was stained with an acridine orange, which is a nucleic acid-specific fluorophore. The same average power of 1.8 mW at the sample and 5.4 MHz repetition rate were used in both cases. Using the shorter excitation wavelength allowed us to obtain a significantly brighter image. Weakly fluorescing features between cell walls at the root's cortex became visible. The histogram in Figure 4(c) more quantitatively shows the increase in the fluorescence signal. It became vividly shifted towards higher PMT voltages at the shorter excitation wavelength. The mean fluorescence signal within the frame increased from 52.8 mV (788.2 nm) to 82.9 mV (761.6 nm), which is 57%. The result is consistent with Ref. [20], showing growing absorption towards shorter wavelengths in the near-infrared.

Next, we tested an unstained sample, a lens tissue. Cellulose paper is known as a source of autofluorescence, and lignosulfonate was speculated as the primary source of the signal [21]. Figures 4(d) and (e) compare the images obtained at two excitation wavelengths, 761.6 nm and 788.2 nm, with the same average power of 2.1 mW (at a 4.9 MHz repetition rate). Using the shorter excitation wavelength allowed us to obtain a brighter image, also visible in the intensity histogram in Fig. 4(f). The mean fluorescence signal within the frame increased from 10.6 mV (788.2 nm) to 7.6 mV (761.6 nm), which is 39%. Cellulose fibers have maximal absorbance between 285 and 395 nm with a peak around 380 nm [22]. Although two-photon absorption spectra might differ from what could be expected by multiplying one-photon absorption spectra by a factor of two [12], excitation wavelength closer to 760 nm increased the TPEF signal.

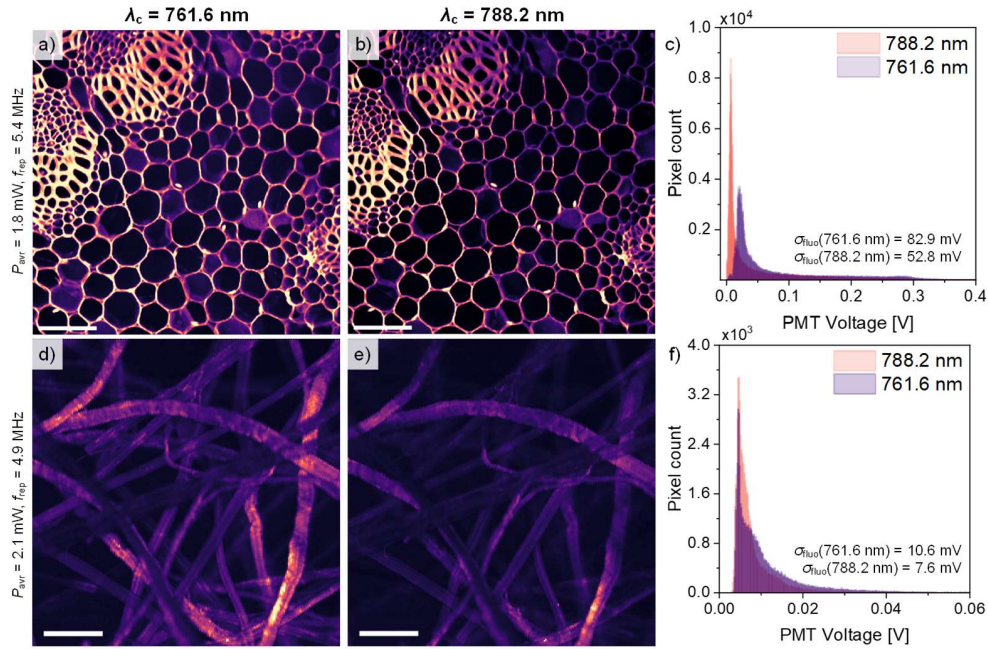

Fig. 4. Adjusting the central wavelength of excitation allows for increasing the brightness of fluorescence images. Images of *convallaria majalis* sample stained with acridine orange obtained at 1.8 mW average power using (a) 761.6 nm and (b) 788.2 nm excitation wavelengths, and (c) histogram of pixel intensity at those two excitation wavelengths ($\sigma_{fluo}$, average pixel intensity within the frame). Images of unstained lens tissue obtained at 2.1 mW average power using (d) 761.6 nm and (e) 788.2 nm excitation wavelengths, and (f) histogram of pixel intensity at those two excitation wavelengths. Scale bar: 50 μm.

In both cases, imaging of stained sample and autofluorescence, adjusting the excitation wavelength to the absorption spectrum of a fluorophore allowed us to obtain improved-quality images.

### 3.3   *Effect of adjustable and reduced pulse repetition rate*

Next, we consider the effect of reducing the pulse repetition rate while maintaining the average excitation power. Such operation effectively translates into increased pulse energy and, simultaneously, increased pulse peak power. Figure 5(a) demonstrates the TPEF signal measured as a function of the average excitation power for five selected pulse repetition rates; a red cast acrylic test sample was used (Thorlabs FSK6). We set our laser to the central wavelength of 782.7 nm and 9.8 MHz, 12.2 MHz, 16.6 MHz, and 24.4 MHz repetition rates. We compared the results with a standard fiber laser working at a 100 MHz repetition rate. The measurement data were fitted with quadratic function ($y = C \cdot x^2$), and parameters of fit functions are given in Table S1. The measured points follow squared dependence on power, as expected for the two-photon effect. However, we note that a saturation effect is visible for lower repetition rates and higher excitation powers; we fitted quadratic functions neglecting points exhibiting this effect. Our data confirmed that reducing the pulse repetition rate at a given average power allows for increasing the TPEF signal. According to Eq. (1), the increase in the fluorescence signal is proportional to the reduction of repetition rate, provided all other parameters are fixed. For example, we obtained a 2.05 times higher value of the C parameter when reducing the pulse repetition rate from 24.4 MHz to 12.2 MHz (a factor of 2 expected due to two times lower repetition rate). Alternatively, a 2.61 increase was observed while the pulse repetition rate was reduced from 24.4 MHz to 9.8 MHz (2.5 factor expected). When compared with a standard fiber laser, using our laser with 9.8 MHz, we obtained a 22.7 times higher value of the C parameter (a factor of 10.2 expected). However, other parameters, such as central wavelength and pulse duration, differed between the standard and our laser, which might explain this difference.

Figure 5(b)-(e) illustrates the same effect with images of the *convallaria majalis* sample. 1.0 mW excitation power was used for all repetition rates, and the same intensity scale was applied for all images. Using a lower repetition rate significantly increased the brightness of obtained fluorescence images. Reducing the pulse repetition rate from 24.4 MHz to 9.8 MHz increased the mean fluorescence signal within the frame by a factor of 2.0 (a factor of 2.5 expected). The reduction from 24.4 MHz to 4.9 MHz gave a 4.7 increase factor (5 expected). When comparing the 4.9 MHz repetition rate with the standard laser, a 36.0 times higher mean fluorescence signal within the frame was obtained. Figure S4 demonstrates the image obtained with the 100 MHz rate with an adjusted intensity scale, showing significantly weaker contrast. Numerical values for all repetition rates are given in Table S2.

As an alternative strategy, reducing the pulse repetition rate could be used to lower the average excitation power needed to produce the same quality image. Figure 6 demonstrates two images, one recorded with our laser with a repetition rate set to 9.8 MHz and one recorded with the standard fiber laser. Both lasers had similar central wavelengths, and average powers were set to obtain images with similar mean fluorescence signals per frame (84 mV and 86 mV, respectively). We used 1.5 mW of average power in the case of our laser and 10.8 mW using the standard laser. The results showed that it is possible to obtain similar-quality TPEF images using approx. 7.2 times lower power by reducing the repetition rate 10.2 times. For such a reduction, a $10.2^{1/2}=3.2$ lower average power needed would be expected, and deviation from this value could result from minor differences in the pulse duration and excitation spectrum of each laser.

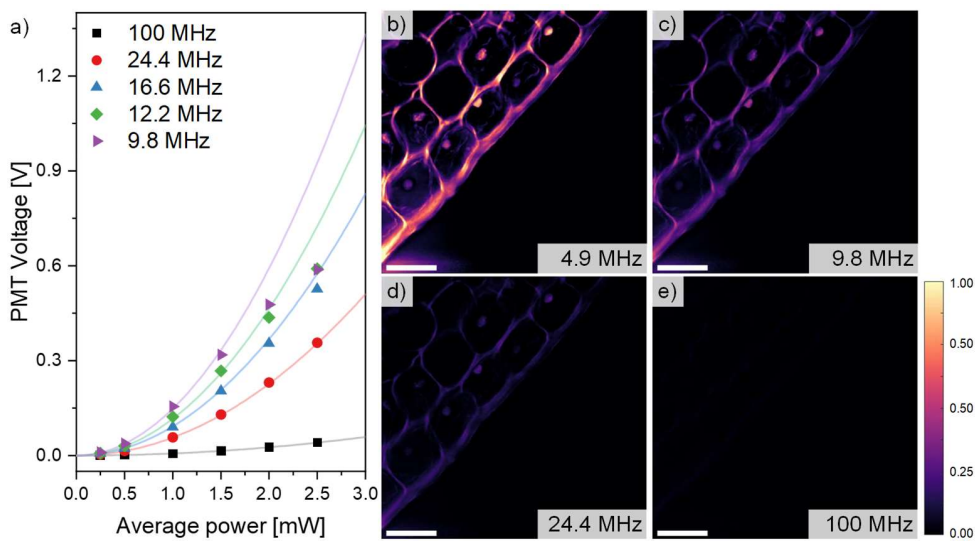

Fig. 5. Reducing the pulse repetition rate while maintaining the average power allows for increasing the brightness of fluorescence images. (a) TPEF signal as a function of average excitation power for selected pulse repetition rates and quadratic fit functions. Images of *convallaria majalis* sample stained with acridine orange obtained at 1.0 mW average power using repetition rate of (b) 4.9 MHz, (c) 9.8 MHz, (d) 24.4 MHz, and (e) 100 MHz (standard fiber laser). Scale bar: 35 μm.

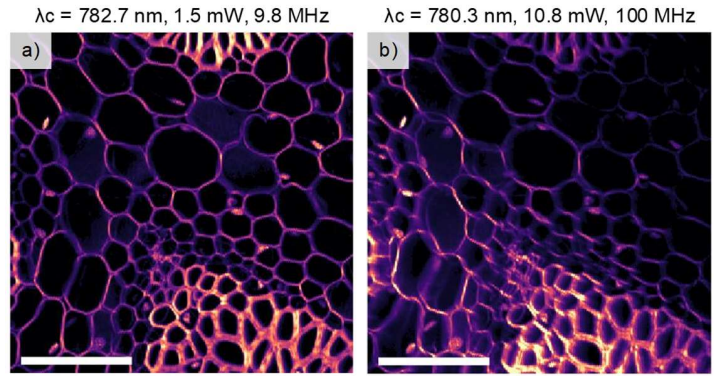

Fig. 6. Reducing the pulse repetition rate allows to produce similar-quality images with reduced average excitation power. Images of the *convallaria majalis* sample stained with acridine orange were obtained using (a) our laser with a 9.8 MHz repetition rate and 1.5 mW power and (b) a standard laser with a 100 MHz repetition rate and 10.8 mW average power. Scale bar: 35 μm.

### 3.4   Effect on penetration depth

Reducing the pulse repetition rate can also improve imaging quality during volumetric imaging at increased depths [23]. We investigated this using an autofluorescent, scattering sample, i.e., two lens tissues stacked one on another (Fig. 7). We imaged this sample using two selected pulse repetition rates, 4.9 and 24.4 MHz, and the same average power of 2.4 mW at the sample plane. The central wavelength was set to 782.7 nm. Supplementary Files 1 and 2 contain 3D objects' obtained at both repetition rates and allow for interactive viewing. Figures 7(a) and (b) show *en face* and side view of the volumetric image (voxel size was 2x2x5 μm) obtained at a lower repetition rate. Cellulose fibers are visible, and the second layer is still noticeable. Figure 7(c) and (d) show two slices obtained at 125 μm and 280 μm depths, respectively. The image obtained from the second layer has a visibly lower contrast but is still informative. On the contrary, a much lower contrast image was obtained with a higher repetition rate, and certain fibers from deeper layers became invisible [Fig. 7(e)]. A side view in Fig. 7(f) shows that the first tissue layer was visible, but the second one became almost unnoticeable, also evident in Figures 7(g) and (h).

### 4.   Discussion

Reducing the pulse repetition rate while maintaining the average excitation power is an extremely efficient yet straightforward way of increasing the brightness of fluorescence images. On the one hand, it could be used to improve the quality of autofluorescence images based on fluorophores with low quantum yield, such as NADH [10], especially *in vivo*. Alternatively, reducing the pulse repetition rate could be used to lower the average power needed during imaging. It could be very beneficial during *in vivo* imaging of pigment-containing tissues, where average laser power is the most severe limiting factor due to thermal-mechanical damage [24]. Examples of such include skin imaging, where melanin heating can cause the formation of cavitation at the epidermal-dermal junction [25], and retinal imaging, where laser-induced heating of the melanin granules was the leading cause of the observed tissue alterations [6]. It has been shown that reducing the pulse repetition rate mitigates these changes. Reducing the pulse repetition rate helps to reduce the cumulative heating effect when the time between the consecutive pulses is longer than the thermal diffusion time of the imaged tissue (e.g., 70 ns in water) [25].

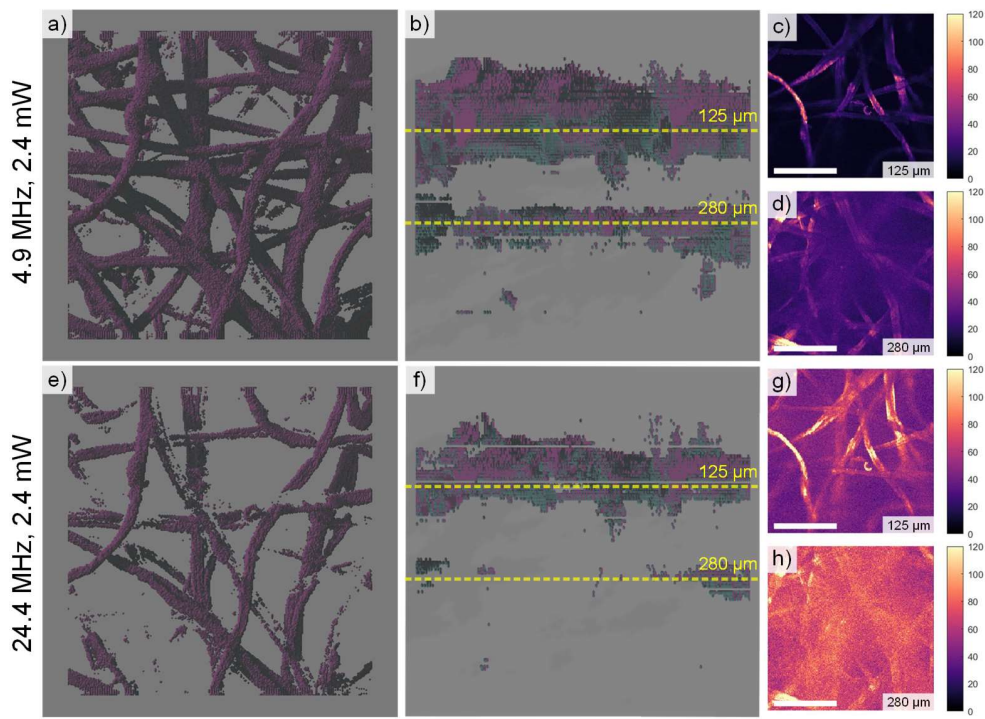

Fig. 7. Reducing the pulse repetition rate while maintaining the average power allows deeper penetration in scattering samples. (a) *En face* and (b) side view of the unstained stack of two lens tissues obtained at 4.9 MHz repetition rate and 2.4 mW of average power. Frames were obtained at (c) 125 μm and (d) 280 μm depths, indicated by the yellow dashed lines. (e) *En face* and (f) side view of the same sample recorded with 24.4 MHz repetition rate and 2.4 mW of average power. Frames were obtained at (g) 125 μm and (h) 280 μm depths. Scale bar: 50 μm.

While this strategy is very efficient, it also might have some drawbacks. First, reducing the pulse repetition rate while maintaining the average power increases the pulse energy. Some indications show that dielectric breakdown may occasionally occur for high peak powers [26]. Secondly, reducing the repetition rate results in a larger number of fluorescence photons per pixel but also in a different distribution of those photons in time. Specifically, more photons are crowded shortly following the excitation pulse. This can cause difficulties in photon counting due to the photodetector dead time and pulse-pair resolution of photon-counting electronics [27]. Lastly, reducing the pulse repetition rate lowers the number of pulses per pixel. In an extreme case, a minimum of one pulse per pixel is required to produce an image. This could be a limiting factor for fast scanning systems and calls for synchronization of scanning protocol with the excitation pulse train. For a quantitative example, imaging with a 1 MHz pulse repetition rate results in a minimum pixel dwell time of 1 μs, translating to 3.8 frames per second for a 512x512 pixel image. The discussed limiting factors, however, highlight the need not for a low repetition rate pulse train but rather a low and adjustable pulse repetition rate. The integrated pulse picker unit allows us to fit this parameter to the particular experimental scenario and imaged sample.

We showed that reducing the pulse repetition rate while maintaining the average power benefits volumetric imaging from within scattering media. Previously, it was achieved by using Ti:sapphire lasers and regenerative amplifiers [28,29]. However, such an approach is complex and expensive, and the reduction of repetition rate is very substantial, limiting imaging speed [9]. The pulse picker approach gives greater flexibility in choosing the pulse repetition rate and is more compact and cost-effective.

## Summary

In summary, by manipulating the temporal and spectral properties of the excitation beam, we have demonstrated a substantial improvement in imaging quality in TPEF microscopy. Alternatively, it was possible to considerably reduce the average power needed to produce the same quality image. Reducing the pulse repetition rate enhances the excitation efficiency by increasing the pulse peak power at maintained average power, and adjusting the wavelength helps to match the absorption spectrum of the sample. To this end, we developed a compact, femtosecond fiber laser with a frequency-doubling stage. The laser operates within the 760 – 800 nm spectral range, producing sub-70 fs pulses with a clean temporal profile and high pulse energy (~1 nJ). The repetition rate could be easily adjusted using an integrated pulse picker unit within the 1 – 25 MHz range without strongly influencing other parameters of the generated pulses. We investigated the effect of reduced pulse repetition and tunable wavelength in TPEF microscopy. We highlighted the need for a low and adjustable pulse repetition rate, which allows us to fit the parameters to the experimental scenario (taking into account imaging speed, sample properties, and system detection parameters). Reducing the repetition rate also benefited the quality of volumetric imaging of scattering samples.

**Funding.** National Centre for Research and Development, Poland (NCBR, LIDER/32/0119/L-11/19/NCBR/2020).

**Acknowledgments.** We thank Aleksander Głuszek and Dr. Arkadiusz Hudzikowski for their help in 3D printing the laser enclosures and designing the heating PCB board, laser diode drivers, and thermoelectric cooling units for amplifiers.

**Disclosures.** The authors declare no competing financial interests.

**Data availability.** Data underlying the results presented in this paper are not publicly available at this time but may be obtained from the authors upon request.

**Supplemental document.** See Supplement 1 for supporting content.